\documentclass[12pt]{article}

\def\be{\begin{equation}}
\def\ee{\end{equation}}
\def\bea{\begin{eqnarray}}
\def\eea{\end{eqnarray}}

\def\lra{\longrightarrow}

\def\3vd{\rangle{\hspace{-0.18em}\longrightarrow}{\hspace{-0.65em}^\mid}}
\def\4vd{\rangle{\hspace{-0.15em}\longrightarrow}}
\def\seagull {\rangle{\hspace{-0.7em}\longrightarrow}}
\def\sw{\!\wedge\!}
\def\dis{\displaystyle}
\def\sl{\hspace{-0.83em}\diagup}
\def\jsl{J\hspace{-0.78em}\diagup}
\def\esl{\epsilon\hspace{-0.4em}/}

\def\kk{\kappa}
\def\pl{\partial^{^{\hspace{-0.23cm}\leftharpoonup}}}
\def\pr{\partial^{^{\hspace{-0.23cm}\rightharpoonup}}}

\usepackage{graphicx}
\usepackage{bm}
\usepackage{amsmath}
\usepackage{amssymb}

\begin{document}

\begin{flushright}
hep-th/0607141
\end{flushright}

\pagestyle{plain}

\begin{center}
\vspace{2.5cm}

{\Large {\bf Recursive Relations For Processes With $n$ Photons Of Noncommutative QED}}

\vspace{1cm}

Abolfazl Jafari

\vspace{.5cm}

{\it Institute for Advanced Studies in Basic Sciences (IASBS),\\
P. O. Box 45195, Zanjan 1159, Iran}

\vspace{.3cm}

\texttt{jabolfazl@iasbs.ac.ir}

\vskip 1.5 cm
\end{center}

\begin{abstract}
The recursion relations are derived for multi-photon processes of noncommutative QED.
The relations concern purely photonic processes as well as the processes with two fermions involved,
both for arbitrary number of photons at tree level. It is shown that despite of the dependence
of noncommutative vertices on momentum, in contrast to momentum-independent color factors of QCD,
the recursion relation method can be employed for multi-photon processes of noncommutative QED.
\end{abstract}



\newpage

\section{Introduction}
The recursion relation method was introduced to study the multi-particle processes in the ``jets" which are produced in
hadron-colliders. The underlying observation that makes this method practical, first made in \cite{parke} and then proved for
general case \cite{giele}, is that the result of perturbative calculation of a gauge theory can be expressed in an
unexpectedly simple and compact form. In the earlier version of the method, the recursion relations are derived for the
multi-gluon current, the so-called gluonic recursion relations \cite{giele}. For the QCD case, as a non-Abelian gauge theory,
this machinery has been developed to give the helicity amplitudes for the processes involving arbitrary number of gluons with
special helicity configurations \cite{giele}; see also \cite{kleiss, kosower, dixon, stir}. Also, apart from
practical point of view, the recursion relations are used to prove
certain properties of amplitudes \cite{giele}. Recently this
technique has been developed to systematic calculation of
amplitudes based on the so-called MHV-rules introduced in
\cite{csw,wu, ozeren,anton}, together with the new recursion relations
now among amplitudes \cite{bcf}, in which amplitudes are
constructed from a new set of building blocks - Maximum Helicity
Violating (MHV) amplitudes - which themselves represent groups of
Feynman diagrams corresponding to particular external helicity
configurations \cite{witten}.

On the other hand, in the last years a great interest has been appeared to study field theories on spaces
whose coordinates do not commute. These spaces, as well as the field theories
defined on them, are known under the names of noncommutative spaces and theories \cite{9908142}; see \cite{reviewnc} as review.
In contrast to U(1) gauge theory on ordinary space-time, as we briefly review in next section,
noncommutative version of theory is involved by direct interactions between photons.
Interestingly one finds the situation very reminiscent to
that of non-Abelian gauge theories, and then the question is whether the techniques
developed for non-Abelian theory purposes can be used for noncommutative QED case too.
In particular, the same question may arise for the recursive relation techniques.
In this contribution we present the recursion relations for photons of noncommutative QED
in the sense of \cite{giele}. This is in fact the first step to employ the recursion relations method
for noncommutative QED case. The recursion method, both in the form of its earlier version \cite{stir} and
recent one \cite{ozeren,anton}, has been already considered for the case of ordinary QED.
As we will see the general structure of these relations for noncommutative QED is similar to QCD's one, though
due to appearance of momentum-dependent factors in vertex functions, instead of momentum-independent
color ones of QCD, a special treatment is needed to manage and reexamine the whole machinery in this case.

The organization of the rest of this work is as follows. In Sec~2 we briefly review some facts
about noncommutative spaces and theories. In Sec~3 we derive the current recursion relations for pure photonic
processes with arbitrary number of photons at tree level. In Sec~4 we present the recursion relations for the processes
in the presence of one pair of fermions, and arbitrary number of photons at tree level. Sec~5 is devoted to conclusion.

\section{Noncommutative Gauge And Dirac Fields}
Noncommutative spacetime coordinates satisfy the commutation relation
\bea
[x^{\mu},x^{\nu}]=i\theta^{\mu\nu}
\eea
\noindent where $\theta^{\mu\nu}$ is a constant real antisymmetric matrix
that parameterizes the noncommutativity of the spacetime.
It is understood that field theories on noncommutative spacetime are defined by actions that
are essentially the same as in ordinary spacetime, with the exception that the products
between fields are replaced by $\star$-product, defined for two functions $f$ and $g$ \cite{reviewnc}
\bea
f(x)\star g(x)&=&f(x)\otimes \,{\rm e}^{\frac{i}{2}\pl_{\mu}\theta^{\mu\nu}\pr_{\nu}}\otimes
g(x)\nonumber\\
&=& f(x)g(x)+\sum_{n=1}^{\infty}(\frac{i}{2})^{n}\frac{1}{n!}
\theta^{\mu_{1}\nu_{1}}\cdots \theta^{\mu_{n}\nu_{n}}
\partial_{\mu_{1}}\cdots \partial_{\mu_{n}}f(x)\partial_{\nu_{1}}\cdots \partial_{\nu_{n}}g(x)
\nonumber\\&&
\eea
\noindent Though $\star$-product itself is not commutative ({\it i.e.}, $f \star g \neq g \star f$)
the following identities make some of calculations easier:
\bea
&~&\int f\star g\;{\rm d}^{n}x=\int g\star f\; {\rm d}^{n}x=\int fg\;{\rm d}^{n}x\nonumber\\
&~&\int f\star g\star h\;{\rm d}^{n}x=\int f (g\star h)\;{\rm d}^{n}x=\int (f\star g) h\;{\rm d}^{n}x\nonumber\\
&~&\int f\star g\star h\;{\rm d}^{n}x=\int h\star f\star g \;{\rm d}^{n}x=\int g\star h\star f\;{\rm d}^{n}x\nonumber
\eea
By the first two ones we see that, in integrands always one of the stars can be removed.
Besides it can be seen that the $\star$-product is associative, {\it i.e.},
$f\star g\star h=(f\star g)\star h= f\star (g\star h)$, and so it is not important which
two should be multiplied firstly.

\subsection{Noncommutative Space-time And U(1) Theory}
The pure gauge field sector of noncommutative U(1) theory is defined by the action
\bea
S_{\rm gauge-field}=-\frac{1}{4}\int {\rm d}^{4}x \;F_{\mu\nu}\star F^{\mu\nu}=-\frac{1}{4}\int {\rm d}^{4}x \;F_{\mu\nu}F^{\mu\nu}
\eea
\noindent in which the field strength $F_{\mu\nu}$ is
\bea\label{field}
F_{\mu\nu}=\partial_{\mu}A_{\nu}(x)-\partial_{\nu}A_{\mu}(x)-ie[A_{\mu}(x),A_{\nu}(x)]_{\star}
\eea
\noindent by definition $[f,g]_{\star}=f\star g-g\star f$. We mention $[x^\mu,x^\nu]_{\star}=i\theta^{\mu\nu}$.
The action above is invariant under local gauge symmetry transformations
\bea\label{trans}
A'_{\mu}(x)\!=\! U\star A_{\mu}(x)\star U^{-1}+\frac{i}{e}U\star
\partial _{\mu}U^{-1}
\eea
\noindent in which $U=U(x)$ is the $\star$-phase, defined by a function $\lambda(x)$ via the $\star$-exponential:
\bea\label{starphase}
&~&U(x)=\exp_{\star}(i\lambda)=1+i\lambda-\frac{1}{2}\lambda\star\lambda+\cdots,\\
&~&U\star U^{-1}=U^{-1}\star U=1
\eea
\noindent in which $U^{-1}=\exp_{\star}(-i\lambda)$. Under above transformation, the
field strength transforms as
\bea
F_{\mu\nu}&\lra& F^{\prime}_{\mu\nu}=U\star F_{\mu\nu}\star U^{-1}
\eea
\noindent We mention that the transformations of gauge field as well as the field
strength look like those of non-Abelian gauge theories. Besides
we see that the action contains terms which are responsible for
interaction between the gauge particles, again as the situation we
have in non-Abelian gauge theories. We see how the
noncommutativity of coordinates induces properties on fields and
their transformations, as if they were belong to a non-Abelian
theory; the subject that how the characters of coordinates and
fields may be related to each other is discussed in \cite{fath}.

\subsection{Feynman Rules Of Noncommutative QED}
\begin{figure}[t]
\begin{center}
\includegraphics[width=0.95\columnwidth]{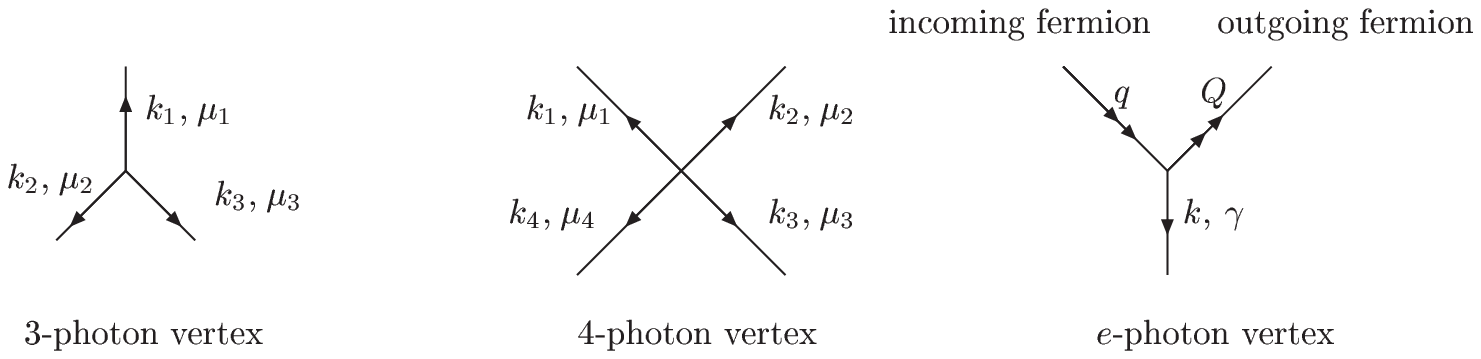}
\caption{Fermion-, 3-, and 4-~photon vertices of noncommutative U(1) theory.}
\label{feyrule}
\end{center}
\end{figure}

Here we present the Feynman rules in the pure gauge theory sector
of noncommutative QED. In this work we restrict ourselves to the
spatial noncommutativity, that is assuming $\theta^{0a}=0$, and
$\theta^{ab}\neq 0$, with $a,b =1,2,3$. We have 3- and 4-vertex
functions, given by (see Fig~\ref{feyrule}) \cite{jab3}: \bea
\Gamma^{\mu_{1}\mu_{2}\mu_{3}}(k_{1},k_{2},k_3)&=& -2e\sin(1\sw 2)
[(k_{1}-k_{2})^{\mu_{3}}g^{\mu_{1}\mu_{2}}
+(k_{2}-k_{3})^{\mu_{1}}g^{\mu_{2}\mu_{3}}\nonumber\\&&
+(k_{3}-k_{1})^{\mu_{2}}g^{\mu_{3}\mu_{1}}] \label{feynman1} \eea
\noindent with $k_3=-(k_1+k_2)$, and \bea
\Gamma^{\mu_{1}\mu_{2}\mu_{3}\mu_{4}}(k_{1},k_{2},k_3,k_4)&=&-4ie^{2}\big[
(g^{\mu_{1}\mu_{3}}g^{\mu_{2}\mu_{4}}-g^{\mu_{1}\mu_{4}}g^{\mu_{2}\mu_{3}}
)\sin(1\sw  2)\sin(3\sw  4)\nonumber\\&&
+(g^{\mu_{1}\mu_{4}}g^{\mu_{2}\mu_{3}}-g^{\mu_{1}\mu_{2}}g^{\mu_{3}\mu_{4}}
)\sin(3\sw  1)\sin(2\sw  4)\nonumber\\&&
+(g^{\mu_{1}\mu_{2}}g^{\mu_{3}\mu_{4}}-g^{\mu_{1}\mu_{3}}g^{\mu_{2}\mu_{4}}
)\sin(1\sw  4)\sin(2\sw  3)\big] \label{feynman3} \eea \noindent
with $k_4=-(k_1+k_2+k_3)$, and the replacement
$\frac{1}{2}k_{i}^a\theta_{ab}k_{j}^b\rightarrow i\sw  j$. Note
that in writing the rules we used the convention that all photons'
momenta are out-going, although one can easily check the
following: \bea k_4 \to -k_4 &\Longrightarrow&
\Gamma^{1,2,3,4}_{\mu_{1}\mu_{2}\mu_{3}\mu_{4}}\to
-\Gamma^{1,2,3,4}_{\mu_{1}\mu_{2}\mu_{3}\mu_{4}} \eea \noindent
and $k_4$ becomes $k_1+k_2+k_3$. As one easily recognizes, the
above vertex functions are very similar to those by non-Abelian
gauge theories, with this exception that the color factors of
non-Abelian theories here are replaced by $\sin(\wedge)$'s
factors, which are momentum dependent. The photon-fermion vertex
function with $k$ for photon's momentum and $Q$ and $q$ for
outgoing and incoming fermion's momenta respectively, in
Noncommutative QED is given by (see Fig~\ref{feyrule}) \cite{jab3}
\bea
\Upsilon^{\mu}(k,Q,q)=ie\gamma^{\mu}\exp{(\frac{i}{2}q^{a}\theta_{ab}
Q^{b})}=ie\gamma^{\mu}\exp{(iq\sw\, Q)} \eea
\vspace{-1cm}
\section{Noncommutative Photon Recursion Relation}
As noted, noncommutative QED is involved by self-interaction of photons, making it in this respect very similar
to a non-Abelian gauge theory. Here we present the photon recursion relations for noncommutative QED.
As we will show the general structure of these relations is similar to QCD's one, though due to appearance of momentum-dependent factors in
vertex functions, instead of color ones in QCD case, a special treatment is needed to manage and reexamine the whole machinery in this case.
In this section we shall consider the tree-level diagrams involved solely by photons. Consider the matrix element for the diagram with $n+1$ outgoing photons,
in the case that one of them is off-shell. This quantity, like its counterpart in QCD case, will be called an
$n$-photon current and is denoted by $\widehat{J}_{\xi}(1,2,\cdots ,n)$,
where $\xi$ denotes the vector index of the off-shell photon.
The $(n+1)$-particle amplitude can be obtained from $\widehat{J}_{\xi}(1,2,\cdots ,n)$ by a
suitable contraction with a polarization vector of the last photon, the $(n+1)\,$th~one.

First let us introduce some parts of the notation hereafter are in use. Following the case for
QCD, we also define the un-hatted $J_{\xi}(1,2,\cdots ,n)$, which essentially defined for the same quantity as $\widehat{J}_{\xi}(1,2,\cdots ,n)$,
apart from the $\sin(\wedge)$ and $\cos(\wedge)$ factors; in the QCD case the difference between $J$ and $\widehat{J}$ is coming from the
color factors \cite{giele}. Hence the helicity's content of current is included only in $J_{\xi}(1,2,\cdots ,n)$.
We define the sum of momenta $\kappa(i,j)=k_i+\cdots+k_j$ with $i<j$ \cite{giele}. Also, we use the replacement
$k_i\sw  (k_j+\cdots+k_n)\to i\sw j\cdots n$.

\begin{figure}[t]
\begin{center}
\includegraphics[width=0.15\columnwidth]{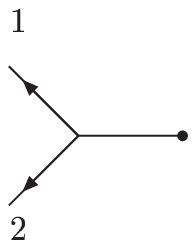}
\caption{Diagram involving two photons and a third off-shell.}
\label{2photon}
\end{center}
\end{figure}
We start with the cases with one, two and three photons. For one
photon we simply define \bea
\widehat{J}_{\xi}(1)=\epsilon_{\xi}=J_{\xi}(1) \eea \noindent
where $\epsilon_{\xi}$ is the polarization vector of the photon,
depending on the helicity and momentum $k_{i}$ of the particle.
Obviously we have $k_{1}\cdot J(1)=0$. For two photons (see
Fig~\ref{2photon}) we use the 3-vertex and introduce a propagator
\bea \widehat{J}_{\xi}(1,2) &=&\frac{2ie\sin(1\sw
2)}{(k_{1}+k_{2})^{2}}\Gamma^{\alpha_{1}\alpha_{2}}_{\xi}
\big(k_{1},k_{2},-(k_{1}+k_{2})\big)
J_{\alpha_{1}}(1)J_{\alpha_{2}}(2)
\nonumber\\
&=&2ie\frac{\sin(1\sw 2)}{(k_{1}+k_{2})^{2}}\big(2k_{2}\cdot J(1)J_{\xi}(2)-2k_{1}\cdot
J(2)J_{\xi}(1)
\nonumber\\&&
+(k_{1}-k_{2})_{\xi}J(1)\cdot J(2)\big)
\eea
As noted earlier we introduce
\bea
J_{\xi}(1,2)
&=&\frac{1}{(k_{1}+k_{2})^{2}}\big(2k_{2}\cdot
J(1)J_{\xi}(2)-2k_{1}\cdot
J(2)J_{\xi}(1)\nonumber\\&&+(k_{1}-k_{2})_{\xi}J(1)\cdot J(2)\big)
\eea
\noindent for which we have
\bea
(k_{1}+k_{2})\cdot
J(1,2)&=&\frac{1}{(k_{1}+k_{2})^{2}}\big(2k_{2}\cdot
J(1)J(2)\cdot k_{1}-2k_{1}\cdot J(2)J(1)\cdot k_{2}
\nonumber\\&&+(k_{1}+k_{2})\cdot(k_{1}-k_{2})_{\xi}J(1)\cdot
J(2)\big)
\nonumber\\
&=&(k_{1}^{2}-k_{2}^{2})J(1)\cdot J(2)=0 \eea \noindent and \bea
J_{\xi}(2,1)&=&\frac{1}{(k_{2}+k_{1})^{2}}\big(2k_{1}\cdot
J(2)J_{\xi}(1)-2k_{2}\cdot
J(2)J_{\xi}(1)\nonumber\\&&+(k_{2}-k_{1})_{\xi}J(2)\cdot J(1)\big)
\nonumber\\ &=&\frac{-1}{(k_{1}+k_{2})^{2}}\big(2k_{2}\cdot
J(1)J_{\xi}(2)-2k_{1}\cdot
J(1)J_{\xi}(2)\nonumber\\&&+(k_{1}-k_{2})_{\xi}J(1)\cdot
J(2)\big)=-J_{\xi}(1,2)
\eea
\noindent So we have both of $(k_{1}+k_{2})\cdot J(1,2)=0$, and $J_{\xi}(1,2)=-J_{\xi}(2,1)$.
By this we can write
\bea
\widehat{J}_{\xi}(1,2)=ie\sum_{P(1,2)}\sin(1\sw  2)J_{\xi}(1,2)
\eea
\noindent with a sum over the permutations of 1 and 2. We also can introduce new notation by using of commutator symbol
\bea
J_{\xi}(1,2)=\frac{1}{(k_{1}+k_{2})^{2}}[J(1),J(2)]_{\xi} \eea
\noindent where \cite{giele}
\bea
[J(1),J(2)]_{\xi}\!\!&=&\!\!\Gamma^{\mu_{1}\mu_{2}}_{\xi} \big(k_{1},k_{2},-(k_{1}+k_{2})\big) J_{\mu_{1}}(1)J_{\mu_{2}}(2)
\nonumber\\
\!\!&=&\!\!\big(2k_{2}\cdot J(1)J_{\xi}(2)-2k_{1}\cdot J(2)J_{\xi}(1)+(k_{1}-k_{2})_{\xi}J(1)\cdot J(2)\big)
\label{93}\eea
\begin{figure}[t]
\begin{center}
\includegraphics[width=0.95\columnwidth]{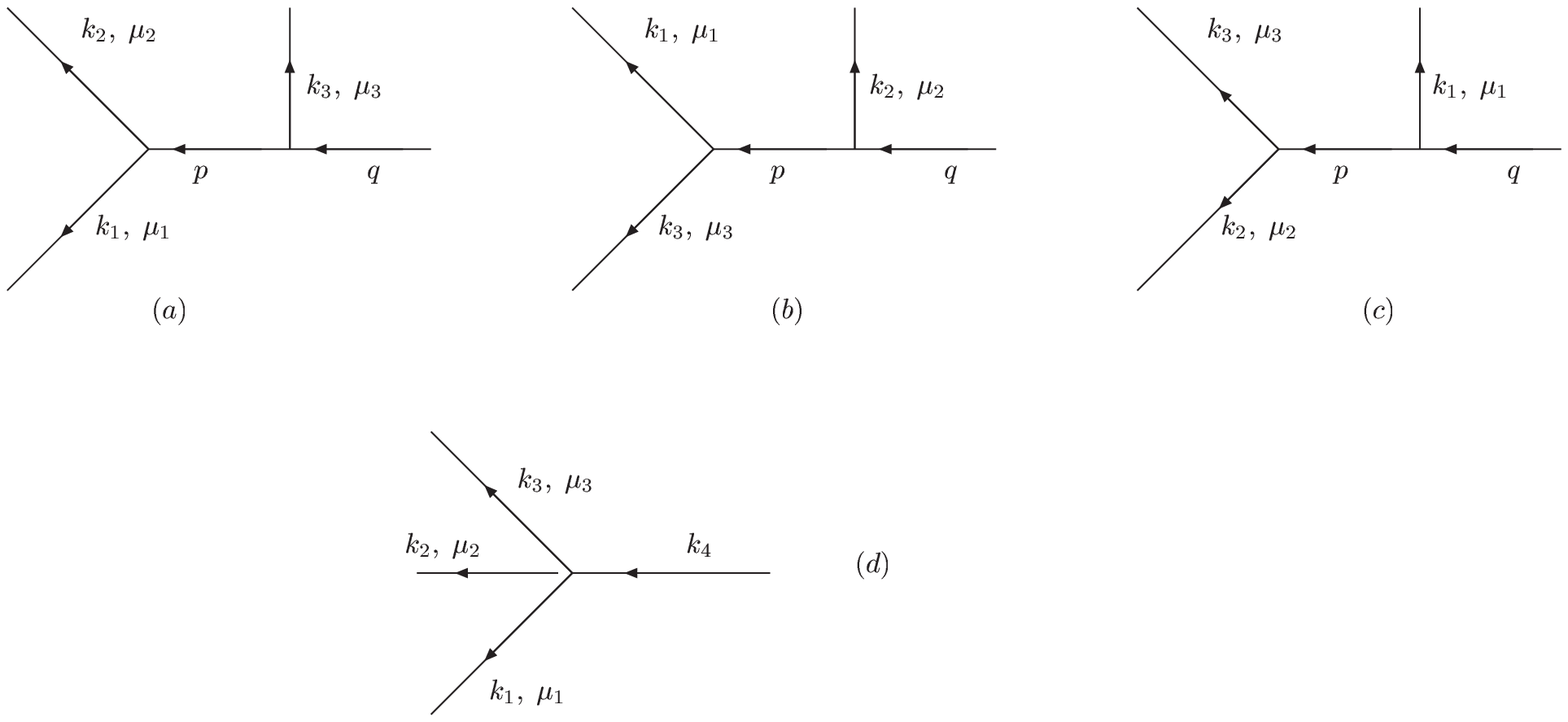}
\caption{Diagrams contributing in current for three photons.}
\label{3photon}
\end{center}
\end{figure}
Now we come to the case with three photons, which is represented by four diagrams (Fig~\ref{3photon}).
The contribution of diagrams $(a)$-$(c)$ are given by
\bea
\Gamma^{\3vd}_\xi=\frac{ (2ie)^2 }{\kappa^2(1,3)}&&\!\!\!\!\!\!\Big(\sin(3\sw 12) \sin (1\sw 2) [J(3),J(1,2)]_\xi\nonumber\\
&&+\sin(1\sw 23) \sin (2\sw 3) [J(1),J(2,3)]_\xi
\nonumber\\
&&+\sin(2\sw 31)\sin (3\sw 1) [J(2),J(3,1)]_\xi\Big)
\eea
\noindent in which, for example
\bea
[J(1),J(2,3)]_{\xi}\!\!&=&\!\!\Gamma^{\mu_{1}\mu_{2}}_{\xi} \big(k_{1},(k_{2}+k_{3}),-(k_{1}+k_{2}+k_{3})\big) J_{\mu_{1}}(1)J_{\mu_{2}}(2,3)\nonumber\\
\!\!&=&\!\!\big(2(k_{2}+k_{3})\cdot J(1)J_{\xi}(2,3)-2k_{1}\cdot J(2,3)J_{\xi}(1)\nonumber\\&&+(k_{1}-(k_{2}+k_{3}))_{\xi}J(1)\cdot J(2,3)\big)
\eea
\noindent By using the identity $2\sin a \sin b = \cos(a-b) - \cos (a+b)$, and the definition $W_\xi(i,jk)=\dis{\frac{1}{\kappa^2(1,3)}}[J(i),J(j,k)]_\xi$,
we find
\bea
\Gamma^{\3vd}_\xi=-2e^2&&\!\!\!\!\!\Big(\cos(1\sw 2+1\sw 3 + 2\sw 3) \,(W(3,12) -W(1,23))_\xi \nonumber\\
&& + \cos(2\sw 3 + 2\sw 1 + 3\sw 1)\, (W(1,23) - W(2,31))_\xi\nonumber\\
&& + \cos(3\sw 1 + 1\sw 2 + 3\sw 2) \,(W(2,31)-W(3, 12))_\xi\Big)
\eea
\noindent $W$'s satisfy the following identities,
\bea
W(i,jk)=-W(jk,i),\;\; W(i,jk)=W(kj,i),\;\; W(i,jk)=-W(i,kj)
\eea
\noindent by which we find
\bea
\Gamma^{\3vd}_\xi=e^2\sum_{P(1,2,3)}\cos(1\sw 2+1\sw 3 + 2\sw 3)\, (W(12,3) +W(1,23))_\xi
\eea
\noindent in which sum is over permutations of $(1, 2,  3)$.

For the seagull  diagram Fig~\ref{3photon}$(d)$ we have
\bea
\Gamma^{\;\;\seagull }_\xi=\frac{-4e^2}{\kappa^2(1,3)}\!\!\!\!&&\Big(\sin (1\sw 2) \sin (3\sw 4) (J(1)\cdot J(3) J_\xi(2)
- J(2)\cdot J(3) J_\xi (1)) \nonumber\\
&&+\sin(2\sw 3)\sin (2\sw 4) (J(1)\cdot J(2) J_\xi(3) -J(1)\cdot J(3) J_\xi(2))\nonumber\\
&&+\sin (3\sw 1) \sin(2\sw 4) (J(2)\cdot J(3) J_\xi(1) - J(2)\cdot J(1) J_\xi (3))\Big)
\nonumber\\&&
\eea
\noindent By definition $I^i_{jk,\xi}=J(i)\cdot(J(j) J_{\xi}(k)-J(k) J_{\xi}(j))$ we have
\bea
\Gamma^{\;\;\seagull }=\frac{-2e^2}{\kappa^2(1,3)}\!\!\!\!&& \Big(\cos(1\sw 2+ 1\sw 3+ 2\sw 3)(I^3_{12}-I^1_{23})_\xi\nonumber\\
&&+ \cos(2\sw 3+ 2\sw 1 + 3\sw 1) (I^1_{23}-I^2_{31})_\xi\nonumber\\
&&+\cos(3\sw 1 + 1\sw 2 + 3\sw 2) (I^2_{31}-I^3_{12})_\xi\Big)
\eea
\noindent and by using the cyclic properties of $I^i_{jk,\xi}$ one finds
\bea
\Gamma^{\;\;\seagull }_\xi=\frac{-e^2}{\kappa^2(1,3)}\sum_{P(1,2,3)} \cos(1\sw 2+ 1\sw 3+ 2\sw 3) (I^3_{12}-I^1_{23})_\xi
\eea
By adding the two contributions of four diagrams we will find:
\bea
\widehat{J}_\xi (1,2,3) &=& e^2 \sum_{P(1,2,3)}\cos(1\sw 2+1\sw 3 + 2\sw 3)\, (W(12,3) +W(1,23))\nonumber\\
&&+\frac{e^2}{\kappa^2(1,3)}\sum_{P(1,2,3)} \cos(1\sw 2+ 1\sw 3+ 2\sw 3) (-I^3_{12}+I^1_{23})_\xi\\
&=&\frac{e^2}{\kappa^2(1,3)}\sum_{P(1,2,3)}\cos(1\sw 2+ 12\sw 3) \, \Big([J(1,2),J(3)]_\xi +[J(1), J(2,3)]_\xi\nonumber\\
&&+\{J(1),J(2),J(3)\}_\xi\Big)
\eea
\noindent in which $\{J(1),J(2),J(3)\}_\xi= (I^1_{23}-I^3_{12})_\xi$. One may write in the form
\bea
\widehat{J}_\xi (1,2,3)=e^{2}\sum_{P(1,2,3)} \cos{(1\!\wedge\! 2+12\!\wedge\! 3)}\;J_{\xi}(1,2,3)
\label{1}
\eea
\noindent in which
\bea
J_{\xi}(1,2,3)=\frac{1}{\kappa^2(1,3)} \Big([J(1,2),J(3)]+[J(1),J(2,3)]+\{J(1),J(2),J(3)\}\Big)_{\xi}
\label{2}
\eea
\noindent One may, as done for $J_{\xi}(1,2)$, show the following identities for the current $J_{\xi}(1,2,3)$:
\bea
&&J(3,2,1)=J(1,2,3)\nonumber\\&&
J(1,2,3)+J(2,3,1)+J(3,1,2)=0\nonumber\\&& \kappa(1,3)\cdot
J(1,2,3)=0
\eea

Now one may try to present the expression for the current
$J_{\xi}$ for the case with $n$ photons. In fact this expression
would be the generalization of (\ref{2}). Here we first suggest an
expression, and do prove it by induction. We start with
\bea
\widehat{J}_{\xi}(1,2,\cdots,n)=(-1)^{n}(ie)^{n-1}\sum_{P(1,\cdots,n)}
\;C(123\cdots n;+\frac{n-3}{2}\pi)J_{\xi}(1,\cdots,n) \label{22}
\eea
\noindent in which
\bea\label{1st;}
C(123\cdots n;+\frac{n-3}{2}\pi)= \cos \Big(&&\!\!\!\!\!\!\!\!\!\! 1\sw 23\cdots n+ 2\sw 34\cdots n+\cdots \nonumber\\
&&+(n-2)\sw(n-1)n+ (n-1)\sw n+\frac{n-3}{2}\pi\Big)
\nonumber\\&&
\eea
\noindent or in the case
\bea\label{2nd;}
C(123\cdots n;)= \cos \Big(&&\!\!\!\!\!\!\!\!\!\! 1\sw 23\cdots n+ 2\sw 34\cdots n+\cdots \nonumber\\
&&+(n-2)\sw(n-1)n+ (n-1)\sw n\Big)
\eea
\noindent Then a generalization of (\ref{2}) for $J_{\xi}$ can be
\bea
J_\xi(1,\cdots,n)&=&\frac{1}{\kappa^2(1,n)}
\Big{(}\sum_{m=1}^{n-1}[J(1,\cdots,m),J(m+1,\cdots,n)]_\xi\nonumber\\&&
+\sum_{m=1}^{n-2}\sum_{k=m+1}^{n-1}\{J(1,\cdots,m),J(m+1,\cdots,k),J(k+1 ,\cdots,n)\}_\xi\Big{)}\nonumber\\
\eea
\noindent This current satisfy the identities
\bea
(a)&& \kappa(1,n)\cdot J(1,\cdots,n)=0\nonumber\\
(b)&& \ J(1,\cdots,n)=(-)^{n-1}J(n,\cdots,1)\nonumber\\
(c)&& \sum_{Cyc(1,\cdots,n)}J(1,\cdots,n)=0 \label{97} \eea
\noindent in which $Cyc(1,\cdots,n)$ is for cyclic permutation on $(1,\cdots,n)$.

One easily can check that for the case $n=1,2,3$ we simply recover the previous expressions. Now we proceed
to prove the suggested expression by induction. Let us assume that the expression is valid for any $m$
with $m<n$. Any current for $n$ photons, just like the case for QCD \cite{giele}, is constructed from considering
a 3-vertex and a 4-vertex with all possible currents attached, here might be called generalized 3-vertex and 4-vertex photons,
represented by $\Gamma_\xi^{1\cdots n,{\4vd}}$ and $\Gamma_\xi^{1\cdots n,\;\,{\seagull }}$. By this the starting point
the expression \cite{giele}
\bea
\widehat{J}_{\xi}(1,\cdots,n)&=&\frac{1}{\kappa^2(1,n)}\sum_{P(1,\cdots,n)}\Big{(}
\sum_{m=1}^{n-1}\frac{1}{2!}\frac{1}{m!}\frac{1}{(n-m)!}\Gamma_\xi^{1\cdots m,m+1\cdots n,{\4vd}} \nonumber\\&& \sum_{m=1}^{n-2}\sum_{k=m+1}^{n-1}
\frac{1}{3!}\frac{1}{m!}\frac{1}{(k-m)!}\frac{1}{(n-k)!}\Gamma_\xi^{1\cdots m,m+1\cdots k,k
+1\cdots n,\;\;{\seagull }}\Big{)}
\label{24}
\eea
\noindent In (\ref{24}) the summation is over all permutations of $n$ photons. In
order to avoid multiple counting factors like $\frac{1}{m!}$ are
introduced, since $\widehat{J}(1,\cdots ,m)$ containing all $m!$
permutation of the particles. Besides we performed breaks in indices $(1,\cdots,n)$ of $\Gamma$'s
according to the momentum flow \cite{giele}
\begin{itemize}
\item In $\Gamma_\xi^{1\cdots n,{\4vd}}$, the momenta of photons $1,\cdots,m$ are going through one of legs of
$\4vd$, and those of other $m+1,\cdots, n$ ones go through the other left leg.
\item In $\Gamma_\xi^{1\cdots n,\;\,{\seagull }}$, the momenta of photons $1,\cdots,m$, $m+1,\cdots,k$, and $k+1,\cdots,n$
are going through each left legs of $\big{\seagull }$.
\end{itemize}
According to the momentum flow above, we would like to introduce a new set of notations for referring the photons belonging
to different legs of $\4vd$ and $\big{\seagull }$. We use $(\alpha,\beta)$ for photons $((1,\cdots,m),(m+1,\cdots,n))$ flowing in left legs of $\big{\4vd}$,
and $(\alpha,\beta,\gamma)$ for photons $((1,\cdots,m),(m+1,\cdots,k),(k+1,\cdots,n))$ flowing in left legs of $\big{\seagull }$.
By these we have the following for $\big{\4vd}$ as examples:
\bea
\sin(\alpha\sw\beta)=\sin\big((1+\cdots+m)\sw ((m+1)+\cdots+n)\big)
\eea
\noindent As further illustration, by $J(\alpha)$ we mean $J(1,\cdots,m)$.
Also we slightly extend the use of symbol ``~;~" in (\ref{1st;}) and (\ref{2nd;}), in the way that here
``~;~" does the same thing in ``~$1\cdots n;$~" here with sets $\alpha$, $\beta$ and $\gamma$. By this we have
\bea
&&\alpha\beta;=\alpha\sw \beta=(1+\cdots+m)\sw ((m+1)+\cdots+n)\nonumber\\
&&\alpha\beta\gamma;=\alpha\sw\beta\gamma+\beta\sw\gamma
\eea
\noindent in which by the convention given in the beginning of this section, $\beta\gamma=\beta+\gamma$.
In occasion, we may use each of $\alpha$, $\beta$ and $\gamma$, in powers and other places, as numbers. As an example,
in $\Gamma^{1\cdots m,m+1\cdots k,k +1\cdots n,\;\;{\seagull }}$, $\alpha$, $\beta$ and $\gamma$ as numbers are simply $m$, $k-m$ and $n-k$, respectively.
By these together, relation like $(b)$ of (\ref{97}) is written as
\bea
J(\beta)\mid_{P(\beta)}\longrightarrow (-)^{\beta-1}J(\beta)
\eea

Now, we can write the generalized 3-vertex function from (\ref{93})
\bea
\Gamma_\xi^{\alpha,\beta\{\4vd\}}&=&ie\sum_{P(\{\alpha\},\{\beta\})}
\sin(\alpha\beta;)[\widehat{J}(\alpha),\widehat{J}(\beta)]_\xi
\nonumber\\
&=&ie\sum_{P(\{\alpha\},\{\beta\})}(-)^{\alpha+\beta}
(ie)^{\alpha+\beta-2}\sum_{P(1,\cdots ,m)}
\sum_{P(m+1,\cdots ,n)} \sin{(\alpha\beta;)}
\nonumber\\
&&C(\alpha;+\frac{m-3}{2}\pi)C(\beta;+\frac{n-m-3}{2}\pi)
[J(\alpha),J(\beta)]_\xi
\nonumber\\
&=&ie\sum_{P(\{\alpha\},\{\beta\})}(-)^{\alpha+\beta}(ie)^{\alpha+\beta-2}
\sum_{P(1,\cdots ,m)}\sum_{P(m+1,\cdots ,n)}
\nonumber\\
&&\Big(\cos(\alpha;-\beta;+\beta\alpha;+\frac{\alpha-3}{2}\pi-\frac{\beta-3}{2}\pi+\frac{\pi}{2})
\nonumber\\
&&+ \cos(\alpha;+\beta;+\beta\alpha;+\frac{\alpha-3}{2}\pi+\frac{\beta-3}{2}\pi-\frac{\pi}{2})
\nonumber\\
&&- \cos(\alpha;-\beta;+\alpha\beta;+\frac{\alpha-3}{2}\pi-\frac{\beta-3}{2}\pi+\frac{\pi}{2})
\nonumber\\
&&- \cos(\alpha;+\beta;+\alpha\beta;+\frac{\alpha-3}{2}\pi+\frac{\beta-3}{2}\pi+\frac{\pi}{2})\Big)
[J(\alpha),J(\beta)]_\xi
\eea
First we notice, by the interchange $\alpha\leftrightarrow\beta$, that
\bea
&&\cos{(\alpha;+\beta;+\beta\alpha;+\frac{\alpha-3}{2}\pi+\frac{\beta-3}{2}\pi-\frac{\pi}{2})}
\times[J(\alpha),J(\beta)]\longrightarrow\nonumber\\&&
-\cos{(\beta;+\alpha; +\alpha\beta;+\frac{\alpha+\beta-3}{2}\pi)}
\times[J(\beta),J(\alpha)]\nonumber\\&& = -\cos{(\beta;+\alpha;
+\alpha\beta;+\frac{\alpha+\beta-3}{2}\pi)}
\times(-[J(\alpha),J(\beta)])\nonumber\\&&
=\cos{(\alpha;+\beta;+\alpha\beta;+\frac{\alpha+\beta-3}{2}\pi)}\times [J(\alpha),J(\beta)]
\eea
\noindent Similarly, by $\alpha\leftrightarrow\beta$ and then doing permutation among $\alpha$'s members, we get
\bea
&&\cos{(\alpha;-\beta;+\beta\alpha;+\frac{\alpha-3}{2}\pi
-\frac{\beta-3}{2}\pi+\frac{\pi}{2})}\times
[J(\alpha),J(\beta)]\longrightarrow
\nonumber\\&&=\cos{(\beta;+\alpha;+\beta\alpha;+\frac{\beta-3}{2}\pi
-\frac{\alpha-3}{2}\pi+\frac{\pi}{2})}\times
[J(\beta),J(\alpha)](-)^{\alpha-1}
\nonumber\\&&=\cos{(\alpha;+\beta;+\alpha\beta;+\frac{\alpha-3}{2}\pi
+\frac{\beta-3}{2}\pi-(\alpha-3)\pi+\frac{\pi}{2})}\times
[J(\beta),J(\alpha)](-)^{\alpha-1} \nonumber\\&&
\eea
\noindent by which, no matter $\alpha$ is even or odd, we get
\bea
&&\cos{(\alpha;+\beta;+\alpha\beta;+\frac{\alpha-3}{2}\pi
+\frac{\beta-3}{2}\pi-(\alpha-3)\pi+\frac{\pi}{2})}\times
[J(\beta),J(\alpha)](-)^{\alpha-1}
\nonumber\\&&=\cos{(\alpha;+\beta;+\alpha\beta;+\frac{\alpha+\beta-3}{2}\pi)}\times [J(\alpha),J(\beta)]
\eea
\noindent Lastly we have
\bea
&&-\cos{(\alpha;-\beta;
+\alpha\beta;+\frac{\alpha-3}{2}\pi
-\frac{\beta-3}{2}\pi+\frac{\pi}{2})}
\times[J(\alpha),J(\beta)]\longrightarrow
\nonumber\\&&=-\cos{(\alpha;+\beta;+\alpha\beta;+\frac{\alpha-3}{2}\pi
-\frac{\beta-3}{2}\pi+\frac{\pi}{2})}\times
[J(\alpha),J(\beta)](-)^{\beta-1} \nonumber\\&&=
\cos{(\alpha;+\beta;+\alpha\beta;+\frac{\alpha+\beta-3}{2}\pi)}\times [J(\alpha),J(\beta)]
\eea
\noindent Using these all the generalized 3-vertex current is given by
\bea
\widehat{J}(1,\cdots ,n)|_{_{\big{\3vd}}}= (-)^{n}(ie)^{n-1}\sum_{P(1,\cdots ,n)}C(1\cdots n;
+\frac{n-3}{2}\pi)J(1,\cdots ,n)
\eea
\noindent in which
\bea
J(1,\cdots ,n)=\frac{1}{\kappa^2(1,n)} \sum_{m=1}^{n-1}[J(1,\cdots
,m),J(m+1,\cdots ,n)]
\eea

For the 4-vertex term we have
\bea
\Gamma^{\alpha,\beta,\gamma,\{\;\;\seagull \}}_{\xi}&=&-4e^{2}
\Big{(}\sin{(\alpha\beta;)}\sin{(\gamma\sw(\alpha+\beta+\gamma))}
\nonumber\\&&
\times(\widehat{J}(\alpha)\cdot\widehat{J}(\gamma)\widehat{J}_{\xi}(\beta)-\widehat{J}(\beta)
\cdot\widehat{J}(\gamma)
\widehat{J}_{\xi}(\alpha))\nonumber\\&&
+\sin{(\beta\gamma;)}\sin{(\alpha\sw(\alpha+\beta+\gamma))}
\nonumber\\&&
\times(\widehat{J}(\alpha)\cdot\widehat{J}(\beta)\widehat{J}_{\xi}(\gamma)-\widehat{J}(\alpha)\cdot\widehat{J}(\gamma)
\widehat{J}_{\xi}(\beta)) \nonumber\\&&
+\sin{(\gamma\alpha;)}\sin{(\beta\sw(\alpha+\beta+\gamma))}
\nonumber\\&&
\times(\widehat{J}(\beta)\cdot\widehat{J}(\gamma)\widehat{J}_{\xi}(\alpha)-\widehat{J}(\beta)\cdot\widehat{J}(\alpha)
\widehat{J}_{\xi}(\gamma))\Big{)}
\eea
\noindent By the definition $\widehat{I}^{\alpha}_{\beta\gamma\xi}=\widehat{J}(\alpha)\cdot(\widehat{J}(\beta)
\widehat{J}_{\xi}(\gamma)-\widehat{J}(\gamma)\widehat{J}_{\xi}(\beta))$, and using the relations
\bea
2\sin{(\gamma\sw(\alpha+\beta+\gamma))}\sin{(\alpha\beta;)}&=&\cos(\alpha
\beta;+\alpha\gamma;+\beta\gamma;) \nonumber\\&&
-\cos(\alpha\beta;+\gamma\alpha;+\gamma\beta;) \eea
\noindent  we can write
\bea
\Gamma^{\alpha,\beta,\gamma,\{{\;\;\seagull }\}}_{\xi}=-2e^{2}&&\!\!\!\!\!\!\!\!\!\Big{(}
\cos(\alpha
\beta;+\alpha\gamma;+\beta\gamma;)(\widehat{I}^{\gamma}_{\alpha\beta}-
\widehat{I}^{\alpha}_{\beta\gamma})_{\xi}
\nonumber\\
&&\!\!\!\!\!\!\!\!\!+\cos(\beta \gamma;+\beta\alpha;+\gamma\alpha;)
(\widehat{I}^{\alpha}_{\beta\gamma}-\widehat{I}^{\beta}_{\gamma\alpha})_{\xi}
\nonumber\\
&&\!\!\!\!\!\!\!\!\!+\cos(\gamma
\alpha;+\alpha\beta;+\gamma\beta;)
(\widehat{I}^{\beta}_{\gamma\alpha}-\widehat{I}^{\gamma}_{\alpha\beta})_{\xi}\Big{)}
\eea
\noindent which can be written in a shorten way
\bea
\Gamma^{\alpha,\beta,\gamma,\{{\;\;\seagull }\}}=-e^{2}\sum_{P(\{\alpha\},\{\beta\},\{\gamma\})}
\cos{(\alpha\beta\gamma;)}
(\widehat{I}^{\gamma}_{\alpha\beta}-\widehat{I}^{\alpha}_{\beta\gamma})_{\xi}
\label{94}
\eea
We introduce $\{\widehat{J}(\alpha),\widehat{J}(\beta),\widehat{J}(\gamma)\}=\widehat{I}^{\alpha}_{\beta\gamma}-\widehat{I}^{\gamma}_{\alpha\beta}$,
satisfying
\bea
\{\widehat{J}(\gamma),\widehat{J}(\beta),\widehat{J}(\alpha)\}=\{\widehat{J}(\alpha),\widehat{J}(\beta),\widehat{J}(\gamma)\}
\eea
\noindent So (\ref{94}) becomes
\bea
\Gamma^{\alpha,\beta,\gamma,\{{\;\;\seagull }\}}&=&e^{2}(-)^{\alpha+\beta+\gamma}(ie)^{\alpha+\beta+\gamma-3}
\nonumber\\
&&\sum_{P(1,\cdots ,m)}
\sum_{P(m+1,\cdots ,k)} \sum_{P(k+1,\cdots ,n)}
\sum_{P(\{\alpha\},\{\beta\},\{\gamma\})}C(\alpha\beta\gamma;)
\nonumber\\
&&C(\alpha;(m-3)\frac{\pi}{2})C(\beta;(k-m-3)\frac{\pi}{2})\nonumber\\&&C(\gamma;(n-k-3)\frac{\pi}{2})
\{J(\alpha),J(\beta),J(\gamma)\}
\eea
\noindent leading to
\bea
\Gamma^{\alpha,\beta,\gamma,\{{\;\;\seagull }\}}&=&e^{2}(-)^{\alpha+\beta+\gamma}(ie)^{\alpha+\beta+\gamma-3}
\nonumber\\
&&\sum_{P(1,\cdots ,m)}
\sum_{P(m+1,\cdots ,k)} \sum_{P(k+1,\cdots ,n)}
\sum_{P(\{\alpha\},\{\beta\},\{\gamma\})}
\nonumber\\&&\frac{1}{8}\Big{(}
\cos{(\alpha\beta\gamma;-\alpha;-\beta;-\gamma;-(\frac{n-9}{2}\pi))}\nonumber\\&&
+\cos{(\alpha\beta\gamma;-\alpha;-\beta;+\gamma;+(\frac{n-2k+3}{2}\pi))}\nonumber\\&&
+\cos{(\alpha\beta\gamma;-\alpha;+\beta;+\gamma;+(\frac{n-2m-3}{2}\pi))}\nonumber\\&&
+\cos{(\alpha\beta\gamma;+\alpha;-\beta;-\gamma;-(\frac{n-2m-3}{2}\pi))}\nonumber\\&&
+\cos{(\alpha\beta\gamma;+\alpha;-\beta;+\gamma;+(\frac{n-2k+2m-3}{2}\pi))}\nonumber\\&&
+\cos{(\alpha\beta\gamma;+\alpha;+\beta;-\gamma;+(\frac{2k-n-3}{2}\pi))}\nonumber\\&&
+\cos{(\alpha\beta\gamma;-\alpha;+\beta;-\gamma;-(\frac{n-2k+2m-3}{2}\pi))}\nonumber\\&&
+\cos{(\alpha\beta\gamma;+\alpha;+\beta;+\gamma;+(\frac{n-9}{2}\pi))}\Big{)}
\nonumber\\&&
\times\{J(\alpha),J(\beta),J(\gamma)\}
\eea
\noindent Like the similar way we should in previous case, by proper interchanges among
$\alpha\leftrightarrow\beta\leftrightarrow\gamma\leftrightarrow\alpha$,
permutations among $\alpha$'s, $\beta$'s and $\gamma$'s members, and and use of
$J(\alpha)\mid_{P(\{\alpha\})}=(-)^{\alpha-1}J(\alpha)$, the eight cosine-terms will be equal to
\bea
-\cos{(\alpha\beta\gamma;+\alpha;+\beta;+\gamma;+(\frac{n-3}{2}\pi))}
\times\{J(\alpha),J(\beta),J(\gamma)\}
\eea
\noindent So
\bea
\Gamma^{\alpha,\beta,\gamma,\{{\;\;\seagull }\}}&=&
e^{2}(-)^{\alpha+\beta+\gamma}(ie)^{\alpha+\beta+\gamma-3}\nonumber\\
&&\sum_{P(1,\cdots ,m)}
\sum_{P(m+1,\cdots ,k)} \sum_{P(k+1,\cdots ,n)}\nonumber\\&&
\cos{(\alpha\beta\gamma;+\alpha;+\beta;+\gamma;+(\frac{n-9}{2}\pi))}
\times\{J(\alpha),J(\beta),J(\gamma)\}\nonumber\\
&=&(-)^{\alpha+\beta+\gamma}(ie)^{\alpha+\beta+\gamma-1}\sum_{P(1,\cdots ,m)}
\sum_{P(m+1,\cdots ,k)} \sum_{P(k+1,\cdots ,n)}\nonumber\\&&
C(\alpha\beta\gamma;+\frac{n-3}{2}\pi)
\times\{J(\alpha),J(\beta),J(\gamma)\} \eea
\noindent Finally
\bea
\Gamma^{\{\;\;\seagull \}}=\sum^{n-2}_{m=1}\sum^{n-1}_{k=m+1}
\Gamma^{\alpha,\beta,\gamma,\{{\;\;\seagull }\}}
\eea
\noindent From this, the second term in (\ref{22}) easily follows.

\begin{figure}[t]
\begin{center}
\includegraphics[width=0.8\columnwidth]{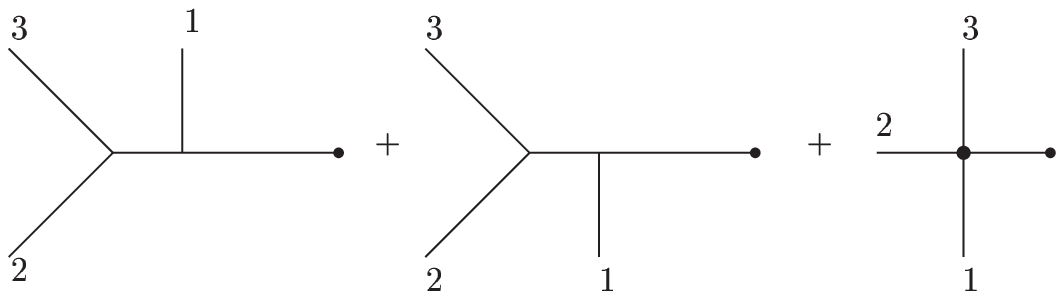}
\caption{Diagrams have a clock-wise orientation for three photons}
\label{fig-4}
\end{center}
\end{figure}
\begin{figure}[t]
\begin{center}
\includegraphics[width=0.9\columnwidth]{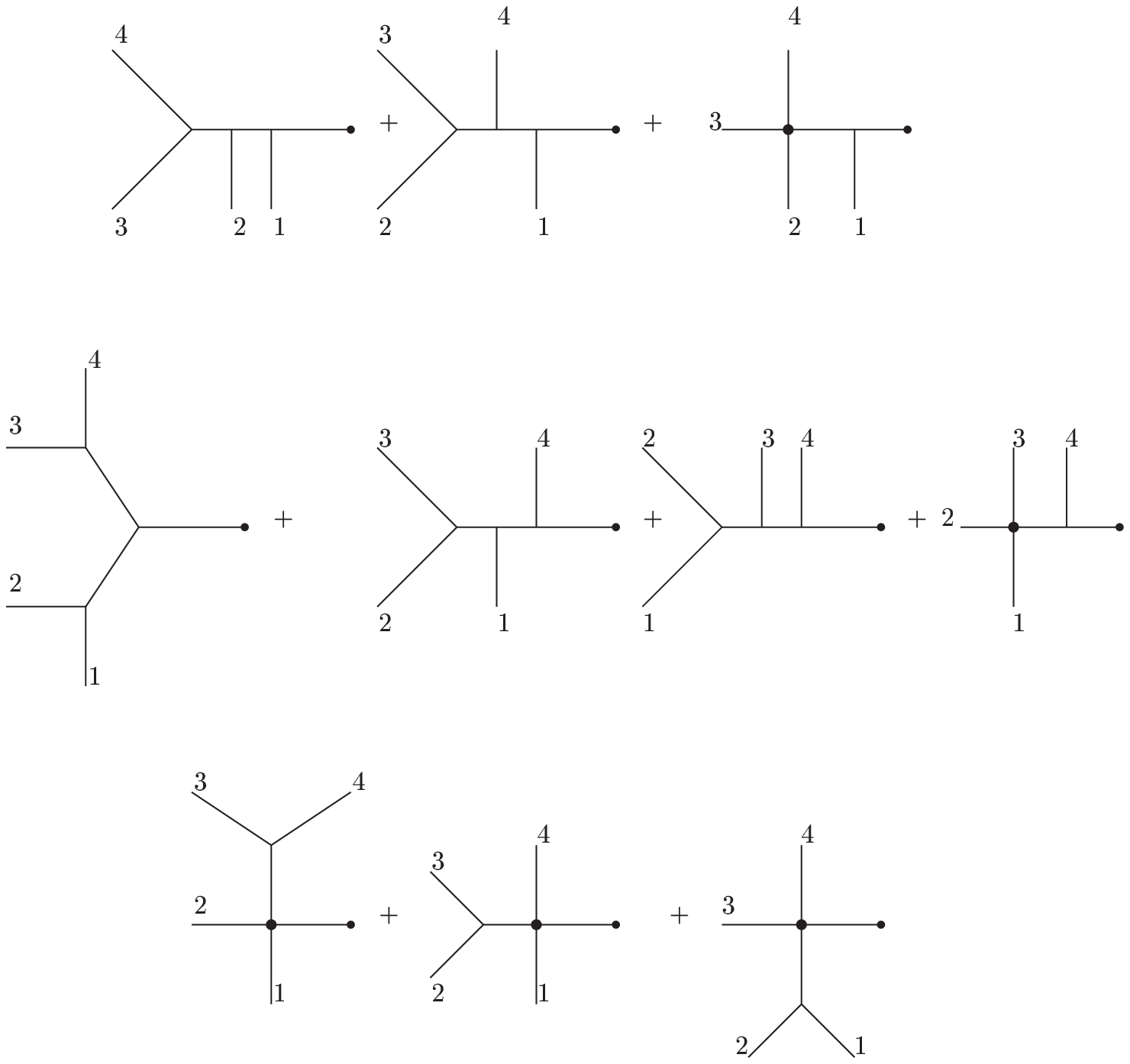}
\caption{Diagrams have a clock-wise orientation for four photon}
\label{fig-5}
\end{center}
\end{figure}

The current $J(1,\cdots ,n)$ has the properties
\bea
J(n,n-1,\cdots ,1)&=&(-)^{n-1}J(1,\cdots ,n)\\
\sum_{Cyc(1,\cdots ,n)}J(1,\cdots ,n)&=&0\\
\kappa(1,n)\cdot J(1,\cdots ,n)&=&0 \eea The properties of $J$ is
like that are proven in \cite{giele}. One may wonder how the
$n$-photon current is affected when we change the gauge of
specific photon, \bea \widetilde{J}(1)=J(1)+\rho k_{1}\label{12}
\eea \noindent We replace $J_{1}$ by $k_{1}$ in recursion relation
and after evaluating the current for a few cases one is led to a
general answer for the current with $J_{1}=k_{1}$ \bea
&&J(1,\cdots ,n)=-\kappa(1,n)\frac{\kappa(1,n)
\cdot\acute{J}(2,\cdots ,n)}{\kappa^2(1,n)}+\acute{J}(2,\cdots
,n)\label{13} \eea \noindent with $\acute{J}(2)=J(2)$, \bea
&&\acute{J}(2,\cdots ,n)=J(1,\cdots
,n)-\sum_{m=2}^{n-1}\frac{\kappa(1,m) \cdot\acute{J}(2,\cdots
,m)}{\kappa^2(1,m)}J(m+1,\cdots ,n) \eea Using induction, this
form can be proven to be correct. The recursion relation include
of clock-wise orientation for the labels $1,2,\cdots ,n$ that is
number of all diagrams occurring in recursion relation; for three
photons there are 3 (Fig~\ref{fig-4}), and for four photons 10
(Fig~\ref{fig-5}) \cite{giele}.

\section{Spinorial recursion relations}
In this section we derive an expression for the matrix
element involving an electron-positron pair and $n$ photons, all
outgoing and the positron off-shell. We introduce the spinorial
current $\widehat{J}(Q,1,2,\cdots ,n)$. In this notation $Q$
stands for the electron's momentum whereas the electron helicity is
suppressed. Moreover $1,2,\cdots ,n$ denote the $n$ photons. The spinor index $c$ of the
off-shell positron is written explicitly.

\begin{figure}[t]
\begin{center}
\includegraphics[width=0.7\columnwidth]{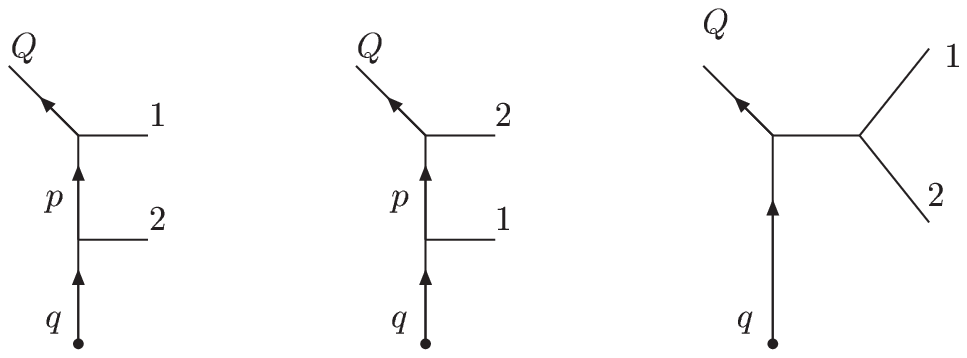}
\caption{Fermion coupled to two photons}
\label{eeph}
\end{center}
\end{figure}

For a single electron and no photon we simply have
\bea
\widehat{J}(Q)=\bar{u}(Q)=J(Q)
\eea
\noindent The one photon spinorial current is
\bea
\widehat{J}(Q,1)=(ie\,{\rm e}^{i1\,\sw\, Q})\bar{u}(Q)\esl_{1}\frac{i}{Q\sl+k\sl_{1}-m_0}
=e\,{\rm e}^{i1\,\sw\, Q}J(Q,1)
\eea
\noindent in which $m_0$ as electron's mass, and with
\bea
J(Q,1)=-J(Q)\jsl(1)\frac{1}{Q\sl+k\sl_{1}-m_0}
\eea
For two photons we have the contribution of three diagrams (see Fig~\ref{eeph}), giving
\bea
\widehat{J}(Q,1,2)&=&(\frac{-e}{Q\sl+k\sl_{1}+k\sl_{2}-m_0})
\Big{(}\,{\rm e}^{i2\,\sw\,   Q1}
\widehat{J}(Q,1)\jsl(2)\nonumber\\&&+ \,{\rm e}^{i1\,\sw\,\,   Q2}
\widehat{J}(Q,2)\jsl(1)+\,{\rm e}^{i12\,\sw\,\,   Q}
J(Q)\sum_{P(1,2)}\widehat{\jsl}(1,2)\Big{)}\nonumber\\&=&
(\frac{-e}{Q\sl+k\sl_{1}+k\sl_{2}-m_0})
\times\sum_{P(1,2)}\Big{(}\,{\rm e}^{i2\,\sw\,   Q1}
\widehat{J}(Q,1)\jsl(2)\nonumber\\&&+\,{\rm e}^{i12\,\sw\,\,   Q}
J(Q)\widehat{\jsl}(1,2)\Big{)}\nonumber\\&=&
(\frac{-e^{2}\,{\rm e}^{i21\,\sw\, Q}}{Q\sl+k\sl_{1}+k\sl_{2}-m_0})
\times\sum_{P(1,2)}\Big{(}\,{\rm e}^{i2\,\sw\,1}
J(Q,1)\jsl(2)\nonumber\\&&+i J(Q)\cos(12;-\frac{1}{2}\pi)
\jsl(1,2)\Big{)}
\eea
\noindent Noting that
\bea
\sum_{P(1,2)}\,{\rm e}^{i12;}
\jsl(1,2)=\sum_{P(1,2)}\,{\rm e}^{-i12;}
\jsl(2,1)\eea
\noindent leads one to
\bea
\widehat{J}(Q,1,2)&=&(\frac{-e^{2}\,{\rm e}^{i21\sw Q}}{Q\sl+k\sl_{1}+k\sl_{2}-m_0})
\nonumber\\&\times&\sum_{P(1,2)}\,{\rm e}^{i2\,\sw\,1}\Big{(}
J(Q,1)\jsl(2)+\frac{i}{2}J(Q)(-\,{\rm e}^{-i\frac{1}{2}\pi}+\,{\rm e}^{i\frac{1}{2}\pi})
\jsl(1,2)\Big{)}
\label{JQ12}
\eea

It is convenient to introduce the abbreviation $K=(k_1,\cdots,k_n)$ for all $k_{i}$'s as photons' momenta. Then,
reminding the action of ``~;~" from previous section, we can write:
\bea
\widehat{J}(Q,1,2)=e^{2}\sum_{P(1,2)}\,{\rm exp}(iKQ;-iK;)J(Q,1,2)
\eea
\noindent with
\bea
\exp(iKQ;-iK;)=\exp\big(i(12\cdots n)\sw Q-i(12\cdots n;)\big)
\eea
\noindent and
\bea
J(Q,1,2)=\frac{-1}{Q\sl+k\sl_{1}+k\sl_{2}-m_0}
(J(Q,1)\jsl(2)+\frac{i}{2}J(Q)(-\,{\rm e}^{-i\frac{1}{2}\pi}+\,{\rm e}^{i\frac{1}{2}\pi})
\jsl(1,2))
\eea
\noindent To express the relation for arbitrary $n$ we need to go one more step, to find
the current for $n=3$ photons, given by
\bea
\widehat{J}(Q,1,2)\widehat{\jsl}(3)+\widehat{J}(Q,1)\widehat{\jsl}(2,3)
+\widehat{J}(Q)\widehat{\jsl}(1,2,3)
=\frac{\mathbb{A}+\mathbb{B}+\mathbb{C}}{Q\sl+\kk\sl(1,3)-m_0}
\eea
\noindent The task is now to find $\mathbb{A}$, $\mathbb{B}$ and $\mathbb{C}$,
\bea
\mathbb{A}&=&-e\sum_{Cyc(1,2,3)}\,{\rm e}^{i3\,\sw\,\, Q12}\widehat{J}(Q,1,2)\jsl(3)
\nonumber\\&=&
-e\sum_{Cyc(1,2,3)}\,{\rm e}^{i3\,\sw\,  Q12}\{ e^{2}\,{\rm e}^{i12\,\sw\,\, Q}
\sum_{P(1,2)}\,{\rm e}^{i2\,\sw\,  1}J(Q,1,2)\}\jsl(3)\nonumber\\&=&
-e^{3}\sum_{P(1,2,3)}\,{\rm e}^{i321\,\sw\,  Q+i32\,\sw\,
12}J(Q,1,2)\jsl(3)\nonumber\\&=& -e^{3}\sum_{P(1,2,3)}\,{\rm exp}(iKQ;-iK;)J(Q,1,2)\jsl(3)
\eea
\noindent Then we have
\bea
\mathbb{B}&=&-e\sum_{Cyc(1,2,3)}\,{\rm e}^{i32\,\sw\,\, Q1}\widehat{J}(Q,1)\widehat{\jsl}(2,3)
\nonumber\\&=&
-e\sum_{Cyc(1,2,3)}\,{\rm e}^{i32\,\sw\,   Q1}\{e\,{\rm e}^{i1\,\sw\,\, Q}J(Q,1)\sum_{P(2,3)}
(+ie\cos(23;-\frac{\pi}{2}))\jsl(2,3)\}
\nonumber\\&=& -ie^{3}\sum_{Cyc(1,2,3)}\,{\rm e}^{i321\,\sw\,  Q+i32\,\sw\,  1}J(Q,1)
\sum_{P(2,3)}(\frac{\,{\rm e}^{i2\,\sw\, 3-i\frac{\pi}{2}}+\,{\rm e}^{-i2\,\sw\,
3+i\frac{\pi}{2}}}{2}) \jsl(2,3) \nonumber\\&=&
-ie^{3}\sum_{P(1,2,3)}\frac{1}{2}\,{\rm e}^{i321\,\sw\,\, Q+i32\,\sw\,
21}J(Q,1) (\,{\rm e}^{i\frac{1}{2}\pi}\jsl(2,3)
+\,{\rm e}^{-i\frac{1}{2}\pi}\jsl(3,2))\nonumber\\&=&
-ie^{3}\sum_{P(1,2,3)}\frac{1}{2}\,{\rm e}^{i321\,\sw\,\, Q+i32\,\sw\,
21}J(Q,1) (\,{\rm e}^{i\frac{1}{2}\pi}
-\,{\rm e}^{-i\frac{1}{2}\pi})\jsl(2,3)
\eea
\noindent Lastly we have
\bea
\mathbb{C}&=&-e\,{\rm e}^{i321\,\sw\,\, Q}\widehat{J}(Q)\widehat{\jsl}(1,2,3)
\nonumber\\&=& -e^{3}\,{\rm e}^{i321\,\sw\,\, Q}
J(Q)\sum_{P(1,2,3)}C (123;)\jsl(1,2,3)
\eea
\noindent Based on the expressions coming from previous section, we have
\bea
&&\sum_{P(1,\cdots ,n)}C(1\cdots n;\frac{n-3}{2}\pi)J_{\xi}(1,\cdots ,n)\nonumber\\
&&=\frac{1}{2}\sum_{P(1,\cdots ,n)}\exp(-iK;)\,(\,{\rm e}^{i\frac{n-3}{2}\pi}J_{\xi}(n,\cdots ,1)
+\,{\rm e}^{-i\frac{n-3}{2}\pi}J_{\xi}(1,\cdots ,n))\nonumber\\&&=
\frac{1}{2}\sum_{P(1,\cdots ,n)}\exp(-iK;)(\,{\rm e}^{i\frac{n-3}{2}\pi}(-)^{n-1}
+\,{\rm e}^{-i\frac{n-3}{2}\pi})J_{\xi}(1,\cdots ,n)
\eea
\noindent by which we finally reach to
\bea
\mathbb{C}=-e^{3}\sum_{P(1,2,3)}\exp(iKQ;-iK;) J(Q)\times\jsl(1,2,3)
\eea

Now we express the general form of spinorial current for the case
with $n$ photons
\bea
\widehat{J}(Q,1,\cdots ,n)=e^{n}\sum_{P(1,\cdots ,n)}\exp(iKQ;-iK;)J(Q,1,\cdots ,n)
\label{14}
\eea
\noindent where
\bea
J(Q,1,\cdots ,n)=\frac{-1}{Q\sl+\kk\sl(1,n)-m_0}
\sum_{m=0}^{n-1}h(n,m)J(Q,1,\cdots ,m)\jsl(m+1,\cdots ,n)
\label{16} 
\eea
\noindent in which
\bea
h(n,m)=\frac{1}{2}(-)^{n-m}(i)^{n-m-1}
(\,{\rm e}^{-i\frac{n-m-3}{2}\pi}
+(-)^{n-m-1}\,{\rm e}^{i\frac{n-m-3}{2}\pi})  \label{17}
\eea
It will be useful to have a spinorial current where the outgoing fermion instead of
the positron is off-shell. The spinorial current with outgoing electron off-shell can be simply obtained.
For one positron it is
\bea
J(P)=v(P)
\eea
\noindent where $P$ is the positron momentum. One easily can check the following replacements
yield the positron currents
\bea
Q&\to&-P\nonumber\\
\frac{-1}{Q\sl+k\sl(1,n)-m_0}&\to&\frac{1}{P\sl+k\sl(1,n)+m_0}\nonumber\\
J(Q,1,\cdots,m)\jsl(m+1,\cdots,n)&\to& \jsl(1,\cdots,m)J(m+1,\cdots,n,P)\nonumber\\
h(n,m)&\to&\tilde{h}(n,m)=\frac{1}{2}(-)^{m}(i)^{m-1}\nonumber\\
&&\times((-)^{m-1}\,{\rm e}^{-i\frac{m-3}{2}\pi} +\,{\rm e}^{i\frac{m-3}{2}\pi})
\eea
\noindent Then the positron currents are given by
\bea
\widehat{J}(1,\cdots ,n,P)=e^{n}\sum_{p(1,\cdots ,n)} \exp(-iKP;+iK;)J(1,\cdots ,n,P)
\eea
\noindent in which
\bea
J(1,\cdots ,n,P)=\frac{1}{P\sl+k\sl(1,n)+m_0}
\sum_{m=1}^{n}\tilde{h}(n,m)\jsl(1,\cdots ,m)J(m+1,\cdots ,n,P) \eea

One can express the relation between two currents by electron and positron by
means of charge conjugation operator ${\cal C}$ \cite{giele}
\bea
{\cal C}v_{\pm}&=&-\bar{u}^{T}_{\pm} \nonumber\\
\bar{u}_{\pm}{\cal C}^{-1}&=&v^{T}_{\pm} \eea \noindent
in
which $v_{\pm}=\frac{1}{2}(1\mp\gamma_{5})v$ and
$u_{\pm}=\frac{1}{2}(1\pm\gamma_{5})u$, and the superscript $T$
for transpose.\\
Following \cite{jab4} we know that ${\cal C}$ is a symmetry of noncommutative theory
when is accompanied with the change $\theta\rightarrow-\theta$, with $\theta$ as the parameter
of noncommutativity.\\
So by means of the action of ${\cal C}$ the currents are related \cite{giele}
\bea {\cal C}
J(1,\cdots ,n,P_{\pm})\mid_{\theta}=(-)^{n-1}J^{T}(P_{\pm},n,\cdots
,1)\mid_{-\theta} \eea \noindent or
\bea J(P_{\pm},1,\cdots ,n)\mid_{\theta}\,{\cal
C}^{-1}=-(-)^{n-1}J^{T}(1,\cdots ,n,P_{\pm})\mid_{-\theta} \eea
\noindent in which $\pm$ denote the helicity of outgoing positron
or electron.

\section{Conclusion}
The $n$-photon recursion relations for purely photonic processes of noncommutative QED are derived.
Also the same relations are presented for processes with one pair of fermions involved.
Although the general structure of these relations for noncommutative QED is similar to QCD's one,
due to appearance of momentum-dependent factors in vertex functions, instead of momentum-independent
color ones of QCD, a special treatment is needed to manage and reexamine the whole machinery in this case.
The relations can be considered as the first step to employ the recursion relation method
for noncommutative QED case.

\vspace{0.5cm}
\textbf{Acknowledgement:} The author is grateful to A. H. Fatollahi for helpful discussions and comments on the manuscript.


\begin{thebibliography}{99}

\bibitem{parke} S. Parke and T. Taylor, Phys. Rev. Lett. {\bf 56} (1986) 2459.

\bibitem{giele} F. A. Berends and W. T. Giele, Nucl. Phys. B {\bf 306} (1988) 759; Nucl. Phys. B {\bf 313}
(1989) 595; Nucl. Phys. B {\bf 294} (1987) 700.

\bibitem{kleiss} F. A. Berends, R. Kleiss, P. de Causmaecker, R. Gastmans and T.~T. Wu,
Phys. Lett. B {\bf 103} (1981) 124; R.~Kleiss and W.~J.~Stirling, Phys. Lett. B {\bf 179} (1986) 2459;
R.~Kleiss and W.~J. Stirling, Nucl. Phys. B {\bf 262} (1985) 235; F.~A.~Berends, R.~Kleiss, P.~de~Causmaecker,
R.~Gastmans, W.~Troost and T.~T.~Wu, Nucl. Phys. B {\bf 206} (1982) 61;  F.~A.~Berends, P.~de~Causmaecker,
R.~Gastmans, R.~Kleiss, W.~Troost and T.~T.~Wu, Nucl. Phys. B {\bf 239} (1984) 382; Nucl. Phys. B {\bf 239} (1984) 395;
Nucl. Phys. B {\bf 264} (1986) 243; Nucl. Phys. B {\bf 264} (1986) 265.

\bibitem{kosower} D.A. Kosower, Phys. Rev. D {\bf 71} (2005) 045007, hep-th/0406175.

\bibitem{dixon} L. J. Dixon, E. W. N. Glover and V. Khoze, JHEP {\bf 0412} (2004)015, hep-th/0411092;
T.~G.~Brithwright, E.~W.~N.~Glover, V.~Khoze and P.~Marquard, JHEP {\bf 0505} (2005) 013, hep-ph/0503063.

\bibitem{stir} R. Kleiss and W. J. Stirling, Phys. Lett. B {\bf 179} (1986) 159.

\bibitem{csw} F. Cachazo, P. Svrcek and E. Witten, JHEP {\bf 0409} (2004) 006, hep-th/0403047.

\bibitem{wu} J.-B. Wu and C.-J. Zhu, JHEP {\bf 0407} (2004) 032, hep-th/0406085;
JHEP {\bf 0409} (2004) 063, hep-th/0406146; L.~J.~Dixon, E.~W.~N.~Glover and V.~Khoze, JHEP {\bf 0412}
(2004) 070, hep-th/0411092.

\bibitem{ozeren} K. J. Ozeren and W. J. Stiriling, JHEP {\bf 0511} (2005) 016, hep-th/0509063.

\bibitem{anton} A. Ilderton, Nucl. Phys. B {\bf 742} (2006) 176, hep-th/0512007.

\bibitem{bcf} R. Britto, F. Cachazo and B. Feng, Nucl. Phys. B {\bf 715} (2005) 499, hep-th/0412308.

\bibitem{witten} E. Witten, Commun. Math. Phys. {\bf 252} (2004) 189-258,
hep-th/0312171; F.~Cachazo and P.~Svrcek, PoS RTN2005 (2005)
004, hep-th/0504194; G.~Georgiou and V.~Khoze, JHEP {\bf 0405} (2004)070, hep-th/0404072.

\bibitem{9908142} N. Seiberg and E. Witten, JHEP {\bf 9909} (1999) 032, hep-th/9908142;
A. Connes, M. R. Douglas, and A. Schwarz,  JHEP {\bf 9802} (1998)  003, hep-th/9711162;
M. R. Douglas and C. Hull, JHEP {\bf 9802} (1998) 008, hep-th/9711165.

\bibitem{reviewnc} M. R. Douglas and N. A. Nekrasov, ``Noncommutative Field Theory,"
Rev. Mod. Phys. {\bf 73} (2001) 977, hep-th/0106048.

\bibitem{fath} A. H. Fatollahi,
Eur. Phys. J. C {\bf 17} (2000) 535, hep-th/0007023;
Phys. Lett. B {\bf 512} (2001) 161, hep-th/0103262;
Eur. Phys. J. C {\bf 21} (Jul. 2001) 717, hep-th/0104210.

\bibitem{jab3} I. F. Riad and M. M. Sheikh-Jabbari, JHEP {\bf 0008} (2000) 045, hep-th/0008132.

\bibitem{jab4} M. M. Sheikh-Jabbari, Phys. Rev. Lett. {\bf 84} (2000) 5265, hep-th/0001167.

\end{thebibliography}
\end{document}